\documentclass[11pt,a4paper]{article}
\usepackage{jheppub}
\usepackage[utf8]{inputenc}

\usepackage{hyperref}
\usepackage{braket}  
\usepackage{subfig}
\usepackage{hhline}
\usepackage{epsfig,amsmath,amssymb,verbatim,mathrsfs,array,layout,textcomp,amssymb,latexsym,slashed,graphicx,booktabs,color,mathtools}
\usepackage{hyperref}
\usepackage{xcolor}
\usepackage[format=plain,textfont=it,justification=justified]{caption}

\newcommand{\beq}{\begin{equation}}
\newcommand{\eeq}{\end{equation}}

\def\beqa{\begin{eqnarray}}
\def\eeqa{\end{eqnarray}}
\def\bea{\begin{eqnarray}}
\def\eea{\end{eqnarray}}

\newcommand{\bv}{\left(\begin{array}{c}}
\newcommand{\ev}{\end{array}\right)}
\newcommand{\bmtwo}{\left(\begin{array}{cc}}
\newcommand{\bmthree}{\left(\begin{array}{ccc}}
\newcommand{\emn}{\end{array}\right)}
\newcommand{\bmtwoc}{\left\{\begin{array}{cc}}
\newcommand{\bmthreec}{\left\{\begin{array}{ccc}}
\newcommand{\emnc}{\end{array}\right\}}
\newcommand{\ba}{\begin{array}}
\newcommand{\ea}{\end{array}}
\newcommand{\be}{\begin{eqnarray}}
\newcommand{\ee}{\end{eqnarray}}

\newcommand{\GeV}{\text{ GeV}}
\newcommand{\TeV}{\text{ TeV}}

\newcommand{\units}[1]{\mathrm{\; #1}}

\def\lsim{\mathrel{\rlap{\lower4pt\hbox{\hskip1pt$\sim$}}
     \raise1pt\hbox{$<$}}}         
\def\gsim{\mathrel{\rlap{\lower4pt\hbox{\hskip1pt$\sim$}}
     \raise1pt\hbox{$>$}}}         

\definecolor{bluDT}{cmyk}{1,0.5,0,0.3}

\usepackage{soul}

\newcommand{\mpar}{M_P}
\newcommand{\mdau}{m_d}

\makeatletter
\gdef\@fpheader{}
\makeatother

\title{dE/dx from boosted long-lived particles}

\author{Gian F.~Giudice,}

\author{Matthew McCullough and}

\author{Daniele Teresi}
\affiliation{CERN, Theoretical Physics Department, Geneva, Switzerland}

\abstract{
At colliders massive long-lived charged particles could be revealed through their anomalously large ionisation energy loss $dE/dx$.  In this paper we explore a class of scenarios in which the LLPs are particularly boosted, owing to production from the decay of a heavy parent resonance.  Such scenarios give rise to unique signatures as compared to traditionally considered $dE/dx$ new-physics benchmarks. We demonstrate that this class of models, unlike traditional new-physics theories, can explain the recently reported excess of events in the $dE/dx$ search by the ATLAS collaboration without conflicting with the determination of $\beta$ from ionisation and time-of-flight measurements.}

\begin{document}
\begin{flushright}
CERN-TH-2022-078
\end{flushright}

\maketitle

\section{Motivation}
In the 1930s Anderson and Neddermeyer noted the observation of unexpected charged long-lived particles (LLPs) `less massive than protons but more penetrating than electrons' \cite{Anderson:1936zz,Neddermeyer:1937md}.  This unexpected discovery of new physics led Rabi to ask `who ordered that?', a question unanswered to this day.   We may infer two lessons for modern particle physics.  The first is that it would not be without precedent if new physics emerged at high energies in the form of charged LLPs.  A second is that a new physics discovery need not conform to any theoretical preconceptions nor answer any particular outstanding theoretical question.  We may, again, be left asking `who ordered that?' for decades to come.

It goes without saying that we should keep our eyes open for new LLPs, wherever we can.  Indeed, there are a wide variety of LLP searches undertaken across a range of energy scales, reflecting the wide range of possibilities that could give rise to them (see e.g.~\cite{Fairbairn:2006gg}).  At the TeV-scale the LHC provides numerous opportunities for LLP discovery.  One is raised by searching for large ionisation energy loss gradients ($dE/dx$) in the tracker of the ATLAS and CMS detectors, see e.g.~\cite{CMS:2011arq,ATLAS:2011ghv,ATLAS:2012urj,CMS:2012wcg,ATLAS:2013cab,CMS:2013czn,ATLAS:2015wsk,ATLAS:2016tbt,CMS:2016kce}. This allows to efficiently distinguish possible new-physics signals from the SM background, which at large $dE/dx$ is small to none, since the known particles with long lifetime are produced relativistically at the LHC.

This can be understood quantitatively by recalling that, for ionising particles with electric charge $Q$ (in multiples of the electron charge) and speed $v$ (with $\beta = v/c$ and $\gamma = 1/\sqrt{1- \beta^2}$), the mean energy loss per distance travelled is given by the Bethe-Bloch relation which, up to density effect corrections, takes the form
\be
-\left\langle \frac{dE}{dx}\right\rangle = 4\pi m_e n_e r_e^2  Q^2 \left(-1 +\frac{2}{\beta^2} \ln \frac{\beta\gamma}{I_e} \right)
 \, . \label{eq:BetheBloch}
\ee
Here $m_e$, $n_e$ and $r_e$ are the electron mass, number density in the medium and classical radius, while $I_e$ is a coefficient related to the mean excitation energy of the medium and the maximum energy transfer in a single electron collision. To adhere with conventions commonly used in high-energy physics~\cite{PDG}, from now on we will refer to $dE/dx$ as the mass stopping power, which is defined as $-\langle dE/dx \rangle /\rho$, where $\rho$ is the mass density of the material. The minus sign in the definition ensures that the mass stopping power $dE/dx$ is positive.

For a particle of unit charge, a large $dE/dx$ signal is expected if the LLP, leaving the ionising track in the Inner Detector, has a relatively small $\beta \lesssim 0.7$. This is the typical working assumption of the collaborations in analysing their data (with exceptions\footnote{These works consider direct production of multi-charged particles. These particles, being relatively slow, would give rise to an ionisation signal much greater than considered here, and a different analysis strategy is typically required.}~\cite{ATLAS:2013cab,CMS:2013czn,CMS:2016kce}). However, a large $dE/dx$ signal is possible also in the alternative assumption of $\emph{fast}$ ionising particles with $Q>1$.  To this end, in this work we develop a strategy to re-interpret the $Q=1$ ATLAS analyses in terms of this hypothesis.

We do not hesitate to add that we are motivated by  the recent  ATLAS announcement of an
excess in the large $dE/dx>2.4 \, \mathrm{MeV g^{-1} cm^2}$ data~\cite{ATLAS:2022pib,ATLAStalk} given by $7$ events in a region with known background of $0.7 \pm 0.4$ events, corresponding to a local (global) significance of 3.6$\sigma$ (3.3$\sigma$), if interpreted as due to metastable gluinos.  Only time will tell whether this excess of events is due to new physics or not. However, since these events lie in a signal-dominated region and the statistical significance will be rapidly tested with new data, it is timely to assess what new physics such events could correspond to.

\begin{figure}[t]
\centering
\includegraphics[width=25em]{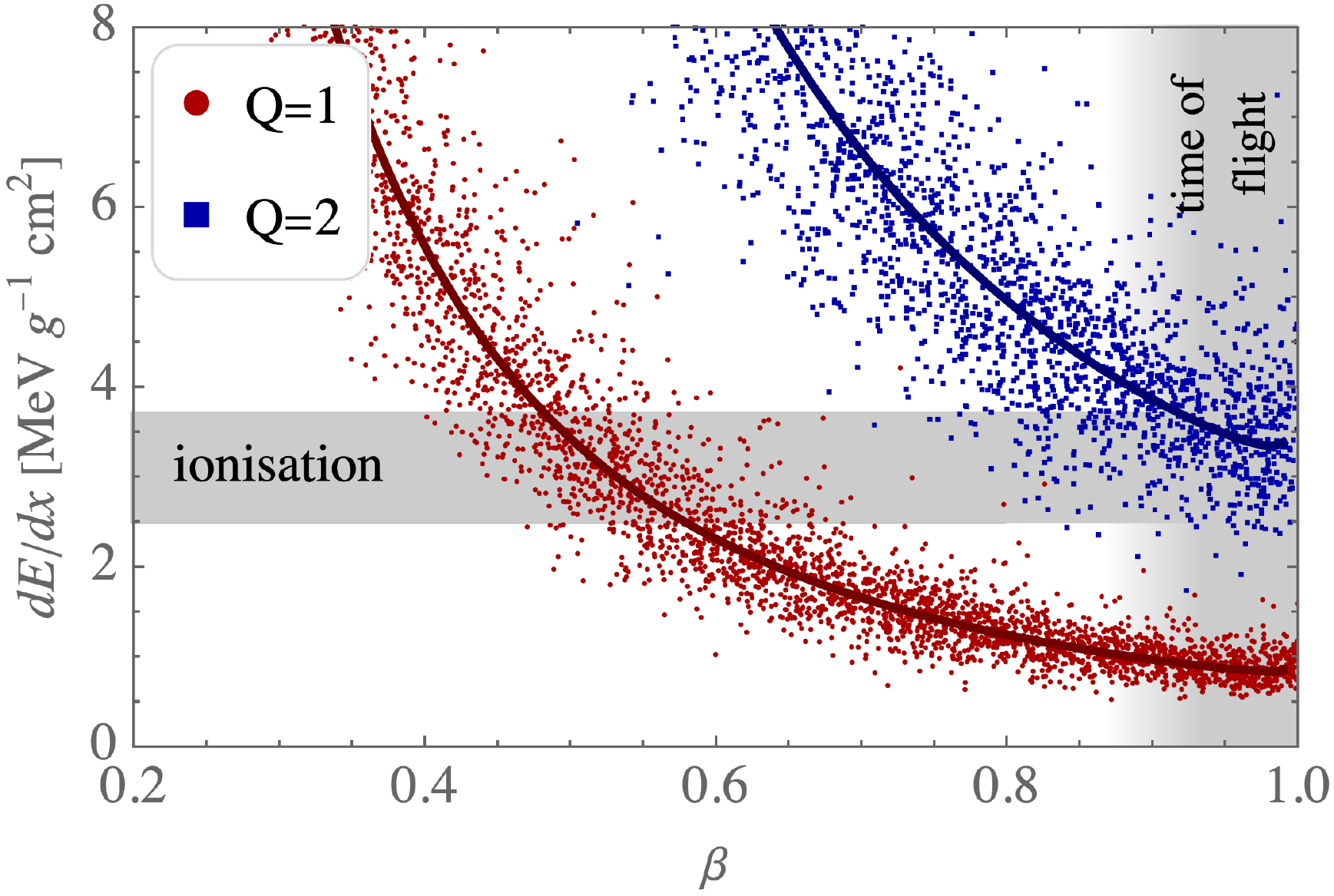} \caption{Distribution of the $dE/dx$ signal as function of $\beta$ for unit and doubly charged particles. The continuous lines denote the most probable values \eqref{eq:dEdxMPV}.  The simulated points are spread around it by the Crystal Ball distribution given in Appendix~\ref{app:calibration}. The grey bands indicate the regions of the ATLAS excess events, in particular their ionisation  $dE/dx$  and $\beta \approx 1$, as suggested by time-of-flight measurements. 
\label{fig:beta}}
\end{figure}

The ATLAS collaboration analyses the excess as being due to relatively slow ($\beta \approx 0.5$) charge $Q=1$ particles with mass in the $1.0$--$2.5 \TeV$ range, considering benchmark models such as $R$-hadrons formed by gluinos in Split-Supersymmetry and metastable charginos or sleptons. However, the time-of-flight determination of $\beta$ instead indicates that all excess events have $\beta \simeq 1$ at $95\%$ confidence level, with uncertainties of about $0.1$ at  2$\sigma$~\cite{ATLAS:2022pib}.  This measurement disfavours any interpretation of the ATLAS excess in terms of conventional LLPs models.  In this work we propose a new scenario of boosted LLPs, which is consistent with observations.  Specifically, we find that the $dE/dx$ excess and the time-of-flight measurements can be explained by boosted ($\beta \approx 1$) particles in the TeV range and with larger electric charge, here $Q=2$ for concreteness, see Fig.~\ref{fig:beta}.  Such boosted states may be produced from the decay of a parent resonance with mass in the range $\mpar \approx 4$--$6 \TeV$, opening the door to a new class of new physics scenarios that may be discoverable at the LHC.

\section{The {\boldmath$dE/dx$} signal for {\boldmath$Q>1$}}\label{sec:dEdx}
We begin by summarising the analysis strategy of the ATLAS collaboration~\cite{ATLAS:2012urj,ATLAS:2015wsk,ATLAS:2016tbt,ATLAS:2022pib}. The most probable value (MPV) of the mass stopping power is calibrated by light SM particles ($\pi^\pm, K^\pm, p$) with $|Q|=1$ \cite{Liang:2014via} and fit using a three-parameter ($c_0,c_1,c_2$) functional form of a phenomenological Bethe-Bloch-like relation
\be
\frac{dE}{dx}\bigg|_{\rm MPV}\!\!\!\!\!\!\!\!\!\!(\beta \gamma) = \frac{1+(\beta\gamma)^2}{\beta\gamma} \Big[ c_0 + c_1 \log_{10}(\beta\gamma) + c_2 \log_{10}^2 (\beta\gamma) \Big] \;.  \label{eq:dEdxMPV}
\ee
For each event with measured momentum $p_{\rm m}$ and ionisation energy loss $dE/dx|_{\rm m}$, an effective mass $m_{dE/dx}$ is obtained by inverting \eqref{eq:dEdxMPV}
\be
\frac{dE}{dx}\bigg|_{\rm MPV}\!\!\!\!\!\!\!\!\!\!(p_{\rm m}/m_{dE/dx}) = \frac{dE}{dx}\bigg|_{\rm m} \;.
\ee
The ATLAS excess is in the region
\be
1.0 \TeV \lesssim m_{dE/dx} \lesssim 2.5 \TeV~~.
\ee
For unit charge particles the effective mass is expected to be close to the physical mass of the particle causing the signal, $m_{dE/dx} \approx m$. However we find that two factors mainly spread $m_{dE/dx}$ around the physical value: the $dE/dx$ distribution around the MPV has a significant width, see Fig.~\ref{fig:beta}, and the momentum resolution in the Inner Detector deteriorates at large values of $p$. This latter factor dominates at the values of interest $p \gtrsim \TeV$.

The unit-charge hypothesis is implicit in the ATLAS analysis, so a direct recast to $Q=2$ is not possible. Thus to analyse the possibility of different charges we develop a different strategy. 

\begin{figure}
\centering
\subfloat[]{\includegraphics[width=25em]{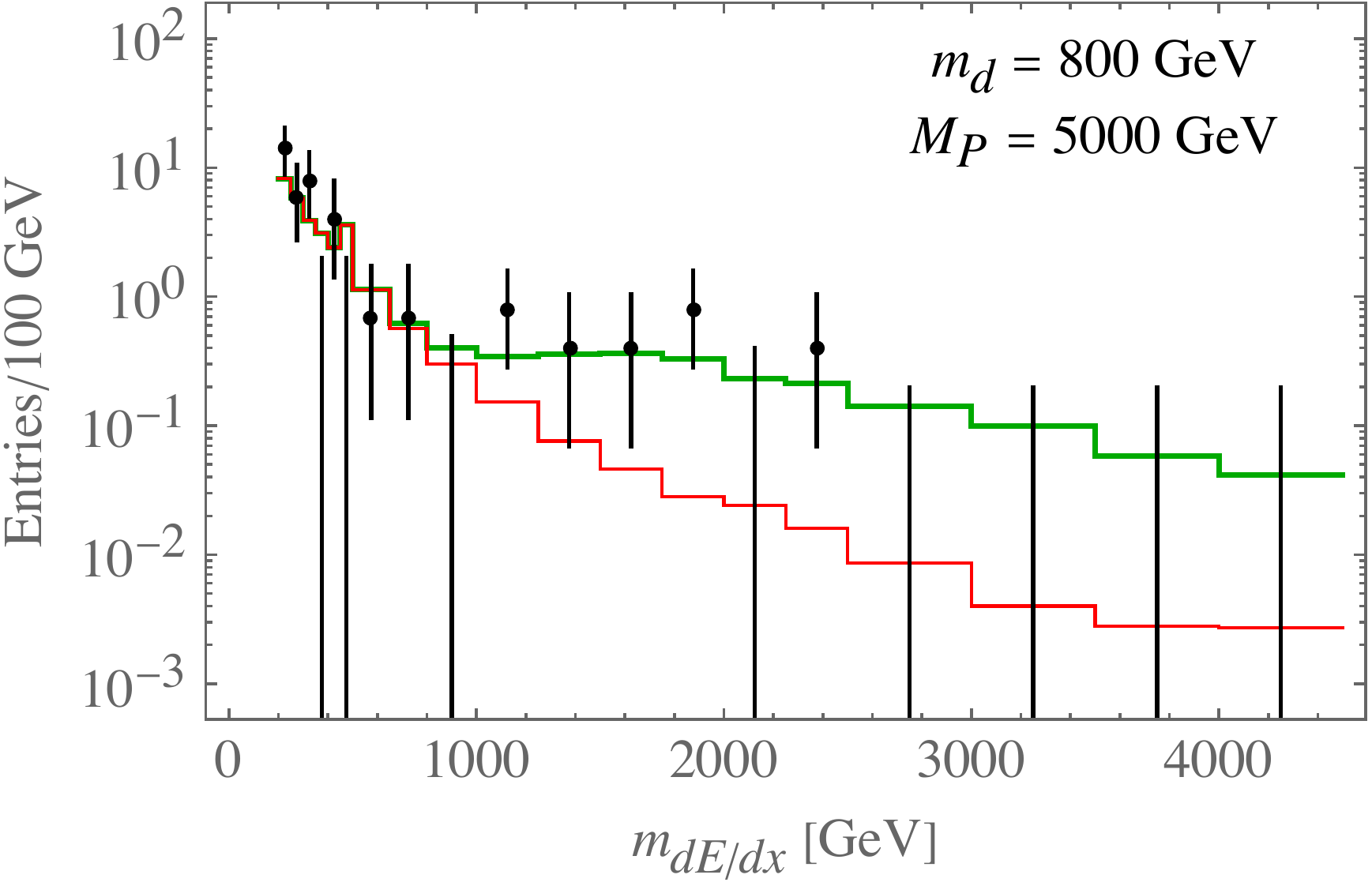}
\label{fig:benchmark_m}}  \vspace{1em} \\
\subfloat[]{\includegraphics[width=25em]{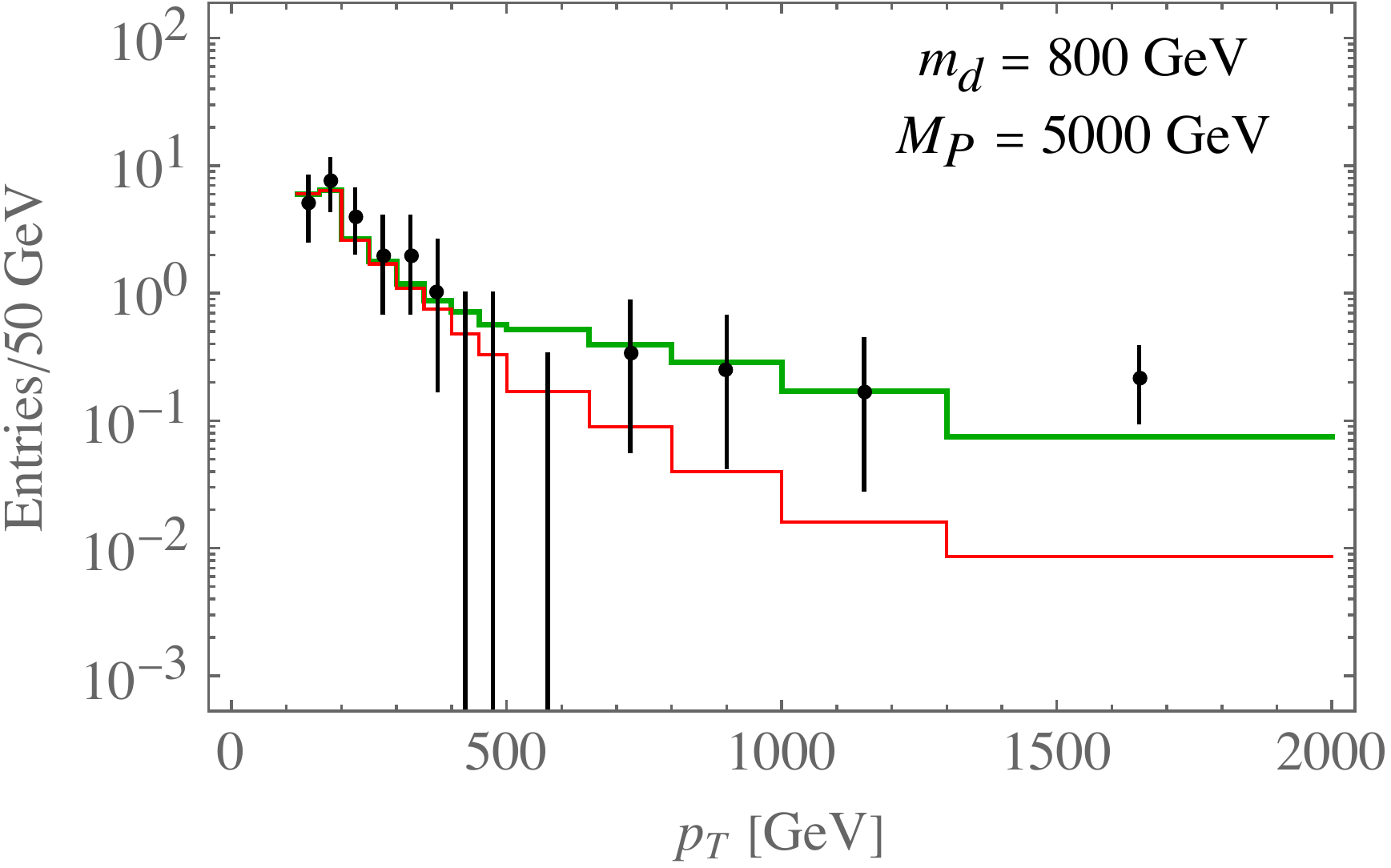} 
\label{fig:benchmark_pT}}  \vspace{1em} \\
\subfloat[]{\includegraphics[width=25em]{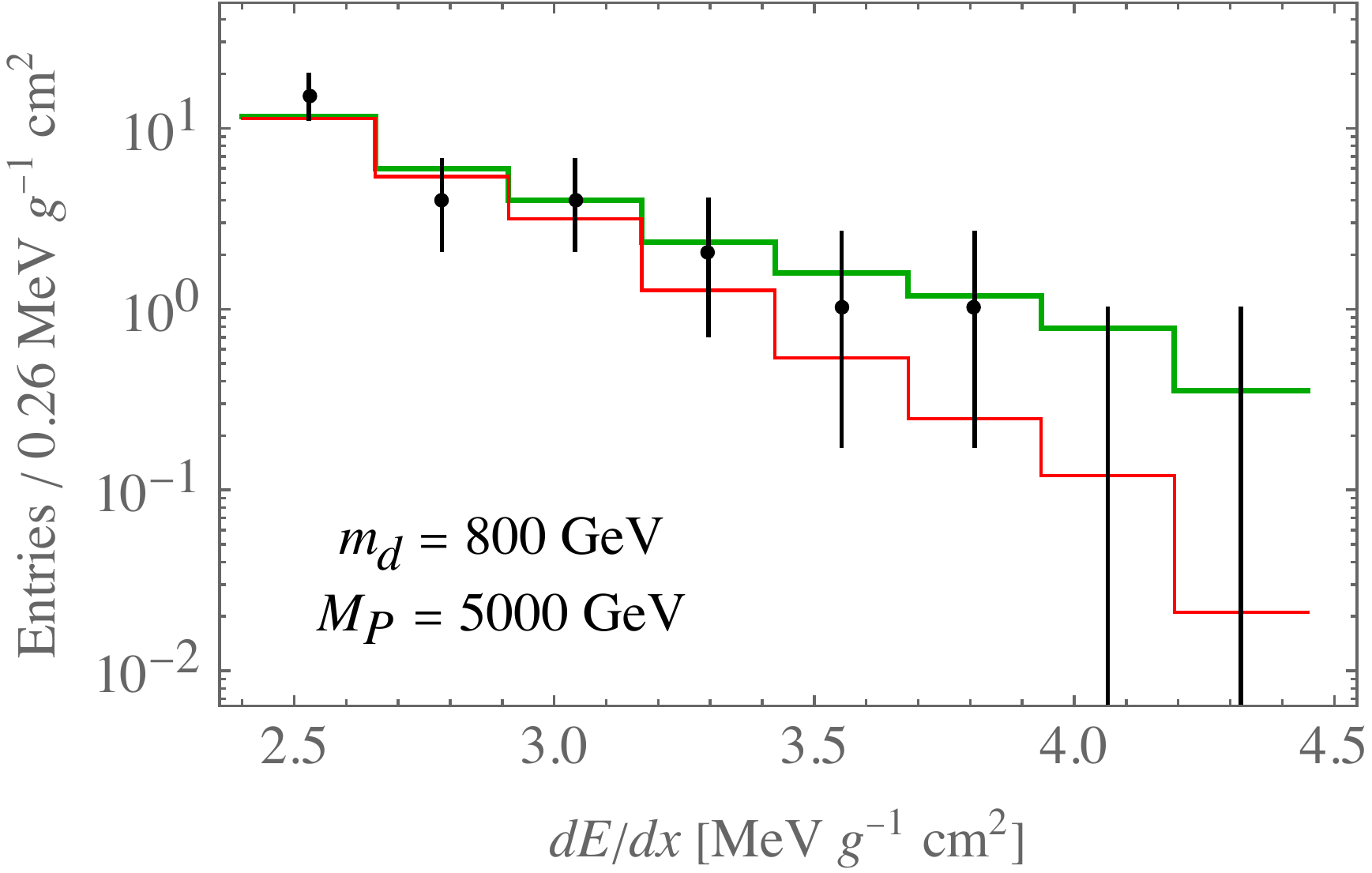}
\label{benchmark_dEdx}}
\caption{{\rm (a)} The effective-mass histogram for the observed data~\cite{ATLAS:2022pib} (black dots), background distribution (red line, taken from~\cite{ATLAS:2022pib}) and background plus $Q=2$ signal model (green line). {\rm (b)} Same for $p_T$.  {\rm (c)} Same for $dE/dx$, with the background extracted from~\cite{ATLAStalk}.} \label{fig:trio}
\end{figure}

We simulate a set of events with true momentum $p$ and ionisation $dE/dx$ given by the distribution around the Bethe-Bloch curve with $Q=2$. This is obtained by the phenomenological relation~\eqref{eq:dEdxMPV}, multiplied by the nominal charge factor $Q^2$.\footnote{This is sufficient at our level of accuracy. Notice that a direct calibration of the Bethe-Bloch curve for $Q>1$ is not available; one could expect an experimental charge factor $Q_{\rm eff}^2$, with $Q_{\rm eff} \approx 2$. We checked that changing $Q_{\rm eff}$ from its nominal value 2 by $\mathcal{O}(10\%)$ does not affect our results significantly.} Then, we take into account that the momentum reconstruction algorithm from the tracker implicitly assumes $Q=1$, since the transverse momentum is obtained as $p_T = Q B \rho$, with $\rho$ being the radius of curvature of the track and $B$ the magnetic field. Therefore, the momentum assigned by the experiments to the event is $p_{\rm rec} = p/2$, half of the real one. In our simulations $p_{\rm rec}$ is then spread by the detector resolution curve reported in~\cite{Benekos:2003xra}. 

Finally, the $Q=1$ ATLAS algorithm is used to generate the $m_{dE/dx}$ histograms, as described at the beginning of this section. Because of the mismatch of charge the effective $m_{dE/dx}$ does not peak around the physical mass $m$ of the particle, but this is not a problem for the analysis, since we find that acceptable signal models are obtained by this procedure, which effectively allows us to interpret possible signals in ATLAS data in terms of $Q=2$ particles.

\section{Results for the ATLAS excess}\label{sec:results}
We now study the excess recently reported by the ATLAS collaboration~\cite{ATLAS:2022pib,ATLAStalk}. To this end, we obtain the Bethe-Bloch curve~\eqref{eq:dEdxMPV} from the calibration data extracted from~\cite{ATLAS:2022pib}, with results reported in Appendix~\ref{app:calibration}. The distribution around the MPV is fitted from the $\pi^\pm, K^\pm, p$ data in~\cite{ATLAS:2022pib} by a one-sided Crystal Ball function. We validate our procedure by reproducing the $Q=1$ signal models considered by the collaboration (see Appendix~\ref{app:validation}).

We consider a model given by a parent resonance with mass $\mpar$, decaying into two metastable $Q=2$ daughter particles with mass $\mdau$, which give rise to the observed excess. Possible explanations for the microscopic origin of this scenario are discussed in the next section. The decay of the parent particle into two daughters is approximated as isotropic in the rest frame and gets longitudinally boosted in the lab frame, working at leading order\footnote{However, we find that the effect of the boost is small, being subdominant compared to the detector momentum resolution.}. We then generate histograms by applying the procedure described in the previous section, with cuts $dE/dx>2.4 \, \mathrm{MeV g^{-1} cm^2}$, $p_T > 120 \GeV$, and perform a profile-likelihood analysis with Poissonian likelihoods.

\begin{figure}[t]
\centering
\includegraphics[width=20em]{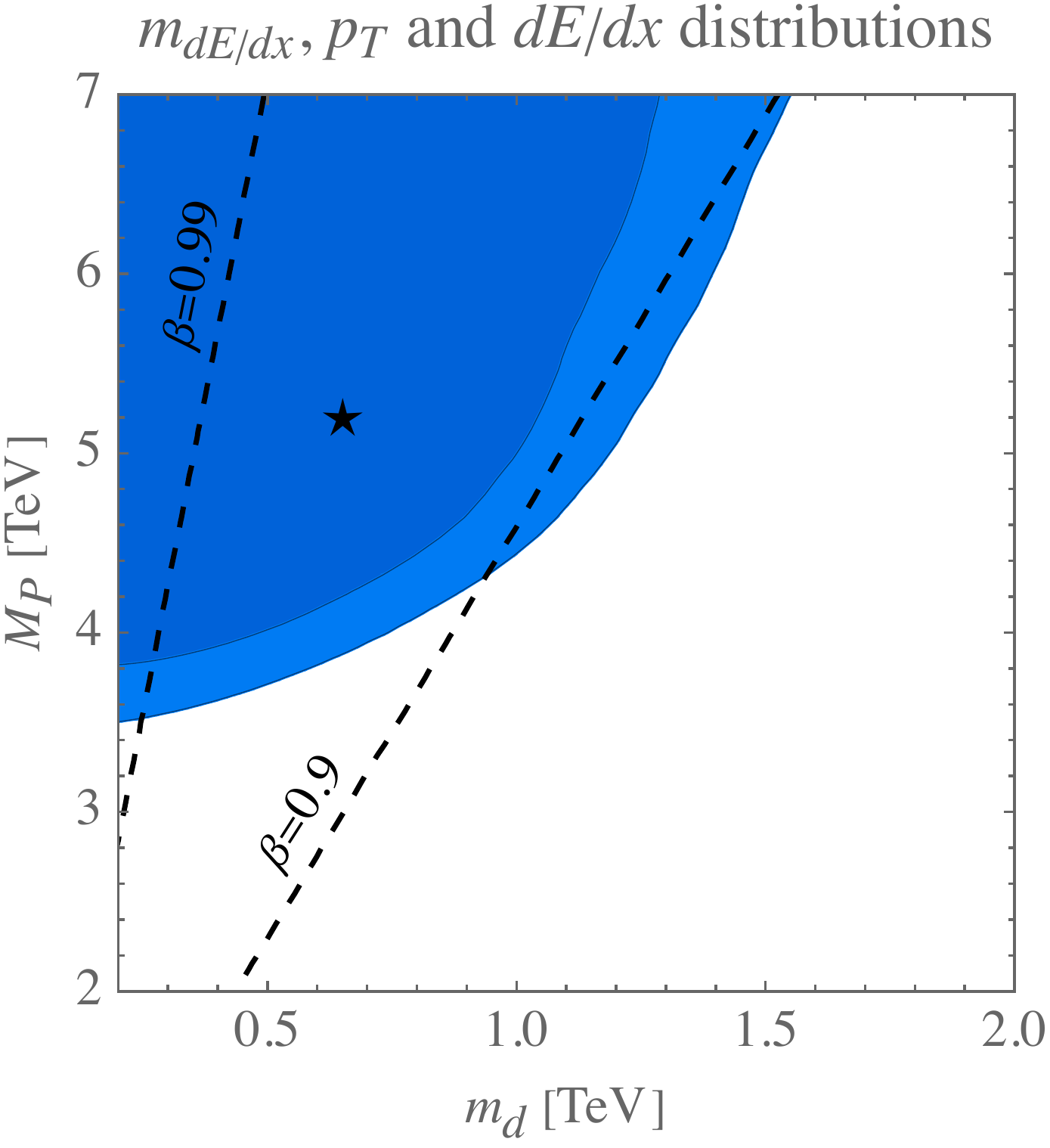} \caption{Profile-likelihood fit of the $m_{dE/dx}$, $p_T$ and $dE/dx$ distributions. Contours show the 1$\sigma$ and 2$\sigma$ preferred regions. The star indicates the best-fit point located at $\mpar = 5.2~ {\rm TeV}$ and $\mdau = 650~{\rm GeV}$. The dashed lines denote the corresponding values of $\beta$ for a decay at rest of the parent resonance.
\label{fig:fit}}
\end{figure}

Our model provides an excellent fit to the excess, as shown in Fig.~\ref{fig:benchmark_m}, with an approximate local significance of about $ 4\sigma$ relative to a background-only scenario\footnote{Following ATLAS, local significances are estimated by fitting the $m_{dE/dx}$ histogram only, using Wilks' theorem.}.
Most importantly, in contrast to the benchmark models considered in~\cite{ATLAS:2022pib}, here the new-physics events have $\beta \approx 1$ by construction, due to the boost resulting from heavy parent decays. We have also checked that the excess is reproduced in the $p_T$ histogram (which is affected by the large $dE/dx$ cut), see Fig.~\ref{fig:benchmark_pT}, and that the $dE/dx$ distribution agrees with data (Fig.~\ref{benchmark_dEdx}).

Finally, we have studied the parameter space that can explain the excess, performing various parameter fits. The results are shown in Fig.~\ref{fig:fit}, obtained fitting all three $m_{dE/dx}$, $p_T$ and $dE/dx$ histograms. Since these are not independent, the confidence intervals are obtained by toy pseudo-experiments, as described in~Appendix~\ref{app:fits}.
All in all, the excess can be explained by charge-two particles in the mass range between hundreds of GeV and the TeV, produced boosted by a parent particle with $\mpar \gtrsim 3.5 \TeV$. 
The $m_{dE/dx}$ distribution alone is well reproduced for $\mpar \gtrsim 2 \TeV$ (see Appendix~\ref{app:fits}), but the inclusion of the $p_T$ information shifts the preferred range of $\mpar$ to larger values. Assuming that the excess persists, the local significance of the best-fit point of our model would reach about $6.2\sigma$ at the end of Run 3 (with $460 \units{fb}^{-1}$ of data).

Figure \ref{fig:fit} describes only the fit to experimental data, but carries no information about the physical production mechanism of the parent particle. As discussed in the next section, realistic models can efficiently produce resonances at the LHC only up to about 6~TeV and this explains why we have limited the vertical axis of Fig.~\ref{fig:fit}. The best-fit point, indicated by a star, lies within the interesting physical mass region.

\section{Microscopic physics} \label{sec:models}
Having demonstrated that the kinematic pattern of heavy resonance production followed by decay to doubly-charged LLPs provides a unique and interesting phenomenological scenario for $dE/dx$ searches, as well as a candidate explanation for the recently observed excess, it naturally follows to briefly explore the microscopic physics that could underlie such a scenario.

Even without specifying the details of the microscopic model, we can derive some general properties of the resonance and its phenomenological consequences from the basic features of our physical setup. We are considering a colourless parent particle $P$ with mass $\mpar$ and spin $J_P$, which can decay into a pair of doubly-charged long-lived daughter particles with branching ratio $B_d \equiv {\rm BR}(P\to dd^c)$. Since $P$ is colour singlet, it can be resonantly produced at the LHC only in the partonic channels $i=gg,q{\bar q}$. We assume that $P$ is coupled to at least one of these possible initial states, with corresponding branching ratio $B_i \equiv {\rm BR}(P\to i)$.

The parent and daughter particles have rather characteristic properties and cannot be immediately embedded in conventional new-physics scenarios. If real, they are likely part of a richer structure and accompanied by other new particles\footnote{Incidentally, ref.~\cite{Akhmedov:2021qmr} made the interesting remark that a doubly-charged long-lived (or stable) scalar particle could combine with light nuclei and catalyse their fusion, even under low-temperature and low-density conditions.}. In this context, it is not surprising that the charge-two daughter is the first particle to be discovered. Indeed, charge-one daughters would have been missed by ATLAS, since they produce a four-times smaller ionisation energy loss (because $dE/dx \propto Q^2$), while neutral daughters do not generate ionisation tracks. 

The total cross section for parent resonant production in $pp$ collisions with centre-of-mass energy $\sqrt{s}$ is, in narrow width approximation ({\it i.e.} $\Gamma_P \ll \mpar$, where $\Gamma_P$ is the $P$ total decay width), 
\beq
\sigma_P =\frac{2J_P+1}{s} \, \frac{\Gamma_P}{\mpar} \sum_i C_i B_i ~,
\eeq
\beq
C_{gg} =\frac{\pi^2}{8} \int_\tau^1 \frac{dx}{x} \, f_g(x) f_g(\tau x) ~,
~~~
C_{q{\bar q}} =\frac{4\pi^2}{9} \int_\tau^1 \frac{dx}{x} \, \left[ f_q(x) f_{\bar q}(\tau x) + f_{\bar q}(x) f_{q}(\tau x) \right]~,
\label{eqCi}
\eeq
where $\tau = \mpar^2/s$ and the parton distribution functions $f_{g,q,{\bar q}}$ are evaluated at $Q^2 = \mpar^2$. The values of $C_i$, for characteristic values of $\mpar$, are tabulated in table~\ref{tab:C}.

\begin{table}[h]
\begin{center}
\begin{tabular}{|c||c|c|c|c|c|c|}
\hline
$\mpar$  {[TeV]} & $C_{gg}$ & $C_{u{\bar u}}$ & $C_{d{\bar d}}$ & $C_{s{\bar s}}$ & $C_{c{\bar c}}$ & $C_{b{\bar b}}$\\
\hline
3 & $3.2 \times 10^{-1}$ & 1.4 & $4.7 \times 10^{-1}$ & $ 1.3\times 10^{-2}$  & $ 4.5\times 10^{-3}$ & $1.7 \times 10^{-3}$\\
4 & $ 2.1 \times 10^{-2}$ & $1.6 \times 10^{-1}$ & $3.2 \times 10^{-2}$ &  $7.3 \times 10^{-4}$  & $ 2.7\times 10^{-4}$ & $9.1 \times 10^{-5}$\\
5 & $1.6 \times 10^{-3}$ & $2.0 \times 10^{-2}$ & $1.8 \times 10^{-3}$ &$4.2 \times 10^{-5}$  & $1.7 \times 10^{-5}$ & $5.6 \times 10^{-6}$ \\
6 & $1.2 \times 10^{-4}$ & $2.1 \times 10^{-3}$ & $8.3 \times 10^{-5}$ & $2.2 \times 10^{-6}$  & $1.1 \times 10^{-6}$ & $3.4 \times 10^{-7}$\\
 \hline
 \end{tabular}
 \caption{The $C_i$ coefficients defined in eq.~\eqref{eqCi}, evaluated using the PDFs MSTW2008NLO~\cite{Martin:2009iq}, for $\sqrt{s}= 13$~TeV and for relevant values of the parent mass $\mpar$.}
 \label{tab:C}
 \end{center}
\end{table}

The number of events with anomalous ionising tracks from fast-moving daughter particles is
\beq
N_{\rm ev} (pp \to P\to dd^c) = {\cal L}\, \epsilon\, \sigma_P \, B_d ~,
\eeq
where $\cal L$ is the integrated luminosity and $\epsilon$ is an efficiency factor, which we take to be 20\%. The requirement of reproducing the 
best-fit signal of 5 events for ${\cal L}=139$~fb$^{-1}$ determines the combination $B_i\, B_d \, \Gamma_P/ \mpar$. In table~\ref{tab:Lam} we show the prediction for a scalar resonance in the gluon channel $(i=gg ,\, J_P=0)$ and for a vector resonance in the quark channel $(i=q{\bar q}  ,\, J_P=1)$, where $B_{q{\bar q}}$ is the branching ratio into a single quark channel taking, for simplicity, a universal value of $B_{q{\bar q}}$ valid for all quark species.\footnote{For the general case of non-universal quark branching ratios, the extracted value of $B_{q{\bar q}}$ has to be interpreted as the weighted average $B_{q{\bar q}}= \sum_i C_i B_i / \sum_i C_i$, where the sum extends over the first five quark species.}

\begin{table}[h]
\begin{center}
\renewcommand{\arraystretch}{1.2}
\begin{tabular}{|c||c|c||c|c|}
\hline
{\footnotesize Resonance} & {\footnotesize Gluon channel} & {\footnotesize Quark channel} & {\footnotesize Scalar resonance} & {\footnotesize Vector resonance} 
 \vspace{-3mm} \\ 
 {\footnotesize mass} &   {\footnotesize$(i=gg ,\, J_P=0)$} & {\footnotesize$(i=q{\bar q}  ,\, J_P=1)$} & {\footnotesize coupled to gluons} & {\footnotesize coupled to quarks} \\
 \hline
$\mpar$ {[TeV]} & $B_{gg}\, B_d \, \Gamma_P/ \mpar$ & $B_{q{\bar q}}\, B_d \, \Gamma_P/ \mpar$ & $\Lambda_P/ \sqrt{B_d}$  {[TeV]} & $g_{Z'}\, |Q_q|\sqrt{B_d} $\\
\hline
3 & $2.4\times 10^{-4}$ & $1.4\times 10^{-5}$ & 12 & 0.013\\
4  & $3.7\times 10^{-3}$ & $1.3\times 10^{-4}$ &  4.0 & 0.041\\
5  & $4.9\times 10^{-2}$ & $1.2\times 10^{-3}$ &  1.4 & 0.12\\
6  & $6.5\times 10^{-1}$ & $1.2\times 10^{-2}$ &  0.4 & 0.39\\
 \hline
 \end{tabular}
 \renewcommand{\arraystretch}{1.0}
 \caption{The combinations $B_i\, B_d \, \Gamma_P/ \mpar$ required to reproduce the ATLAS $dE/dx$ signal, in the case of gluon and quark channels. Also shown are the predictions for the combinations $\Lambda_P / \sqrt{B_d}$ (for the model with scalar resonance coupled to gluons) and  $g_{Z'} |Q_q|\sqrt{B_d} $ (for the model with vector resonance coupled to quarks).}
 \label{tab:Lam}
 \end{center}
\end{table}

The results in table~\ref{tab:Lam} show that, for the gluon channel, the narrow width approximation can hold for $\mpar \lesssim 5$~TeV, but it deteriorates at larger masses where the parent's interactions start becoming non-perturbative. The situation is more favourable for the quark channel, where the narrow width approximation can be satisfied in a broader range of $\mpar$, as long as both branching ratios $B_{q{\bar q}}$ and $B_d$ are not too small. 

A robust and model-independent prediction of our setup with boosted LLPs is an excess of dijet events mediated by resonant parent production. This excess is simply correlated with the anomalous $dE/dx$ events and the ATLAS signal predicts a non-standard contribution to the total dijet cross section at $\sqrt{s} = 13$~TeV
\beq
\sigma_P ({\rm dijet}) = 0.45~{\rm fb} \, \frac{B_{q{\bar q}}}{B_d} ~,
\label{sdijet}
\eeq
where we have included a 50\% efficiency factor and summed over five quark species in the final state, with a universal value of $B_{q{\bar q}}$.\footnote{For the general case of non-universal quark branching ratios, $B_{q{\bar q}}$ in eq.~\eqref{sdijet} has to be interpreted as the average $B_{q{\bar q}}= \sum_i  B_i / 5$, where the sum extends over the first five quark species.}

\begin{figure}[t]
\centering
\includegraphics[width=25em]{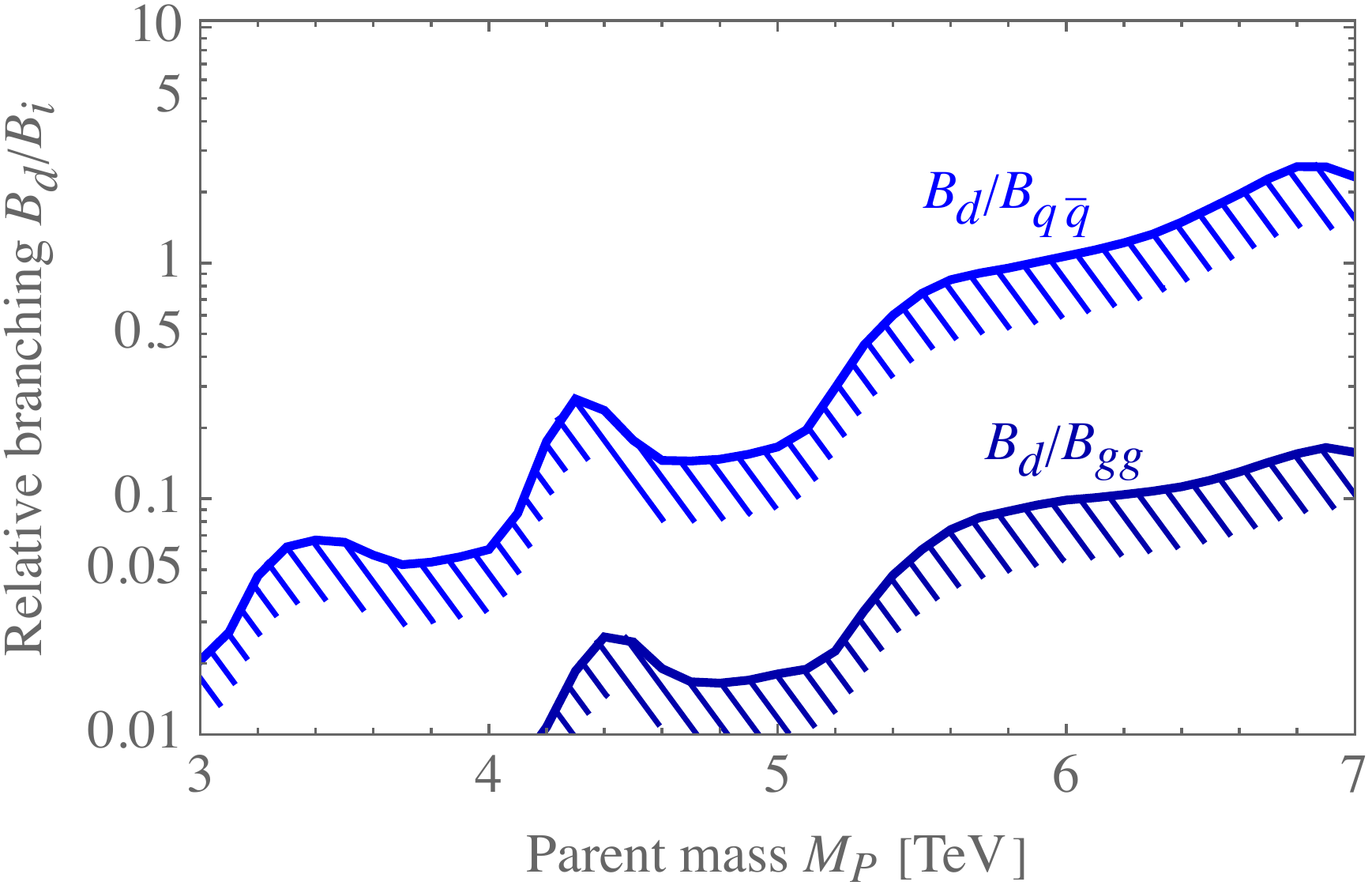} \caption{Lower bounds on the relative branching $B_d/B_i$, both for the gluon ($i=qq$) and quark ($i=q\bar q$) production channels, obtained from the CMS limits on dijet searches~\cite{CMS:2019gwf}. Other parameters have been fixed to reproduce the ATLAS $dE/dx$ signal. 
\label{fig:dijet}}
\end{figure}

Using the LHC limits on resonant dijet production in gluon and quark channels~\cite{CMS:2019gwf}, we can extract the lower bounds on the ratios $B_d/B_{gg}$ and $B_d/B_{q \bar q}$ shown in Fig.~\ref{fig:dijet}. This figure shows that resonant dijet production at present gives only a mild constraint on the boosted LLPs interpretation of the $dE/dx$ excess.

Since the doubly-charged daughters must carry EW quantum numbers,
another generic phenomenological consequence of our setup is an irreducible Drell-Yan production of daughter pairs. This must be subdominant as compared to production via $P$ decay, because the former would give slower $d$ particles with much larger $dE/dx$.  We estimate that for $\mdau \gtrsim 500 \GeV$ this effect is subdominant.

While excess of dijet events and EW pair-production of doubly-charged particles are unescapable and model-independent consequences of our interpretation of the ATLAS $dE/dx$ signal, other experimental signatures (such as dilepton or missing energy from parent decay, or contact interactions from virtual parent or daughter exchange) could be present in specific model realisations, as will be shown in the following.

We can now gain further insight on the microscopic structure of boosted LLPs by illustrating specific examples of models for the parent and daughter particles.

\subsection*{Scalar resonance coupled to gluons}
A simple microscopic model is given by a heavy singlet scalar $P$ coupled to gluons and doubly-charged daughter particles $d$ as
\beq
\frac{\alpha_s}{\Lambda_P} \, P \, G^2_{\mu\nu} + \kappa\, P\, d^c d~,
\eeq
where $\Lambda_P$ is the scale of the dimension-five effective interaction and $\kappa$ is a coupling constant. We do not need to specify the daughter's spin. The model must also include a feeble interaction, possibly described by a higher-dimension effective operator, that allows for $d$ decay, making the daughter metastable.

The parent decay width into gluons is given by
\beq
\frac{\Gamma_{gg}}{\mpar}  = \frac{2 \alpha_s^2 \mpar^2}{\pi \Lambda_P^2}~.
\eeq
The ATLAS $dE/dx$ signal predicts the effective scale of the model, through the combination $\Lambda_P / \sqrt{B_d}$, as shown in table~\ref{tab:Lam} for relevant values of $\mpar$.  

Since the ratio $\mpar/\Lambda_P$ has dimensions of coupling (much like the commonly encountered combination $m_W/v$ in the SM), the model indicates a moderately strongly-coupled UV completion for $\mpar \lesssim 5$~TeV, at least for not too small $B_d$. For larger $\mpar$, the theory enters a strongly-coupled regime and any perturbative control is lost. Therefore, a generic consequence of this model is the likely existence of new coloured states not far beyond the mass scale of $\mpar$, which should not exceed about 5 TeV.

Finally, we remark that our results are unchanged if the parent, instead of being scalar, is a pseudoscalar coupled to $G \widetilde{G}$, since the formul\ae~for the cross section and branching ratio remain the same.  

\subsection*{Vector resonance coupled to quarks}
As an alternative microscopic model, one can take the parent resonance to be a $Z'$ boson of a $U(1)'$ gauge group under which at least the daughter particle and first-generation quarks are charged.\footnote{$Z'$ decays to pairs of unit-charge LLPs were discussed in \cite{Bauer:2009cc}.} Taking the simple case of a vector current with coupling constant $g_{Z'}$, the $Z'$ partial width of the decay into each particle pair $\psi$ is\footnote{The normalisation is chosen such that the gauge interaction of the fermionic current is $g_{Z'} Q_\psi Z'_\mu {\bar \psi} \gamma^\mu \psi$.}
\beq
\frac{\Gamma (Z'\to {\bar \psi}\psi)}{ M_{Z'}} = \frac{{\cal N}_\psi \,  Q_\psi^2 \, g_{Z'}^2}{12\pi} ~,
\eeq
where $Q_\psi$ is the $\psi$ charge under the new $U(1)'$ gauge group and ${\cal N}_\psi$ is the number of effective species. Quarks correspond to ${\cal N}_q =3$, while the daughter particle  gives
\beq
{\cal N}_d = \left\{ 
\begin{array}{ll}
\frac{N_d \, \beta (3-\beta^2)}{2} &{\rm for}~J_d=1/2 \\
\frac{N_d \, \beta^3}{2} &{\rm for}~J_d=0
\end{array} \right.
 ~,~~~ \beta = \sqrt{1-\frac{4\mdau^2}{M_{Z'}^2}} ~,
\eeq
where $N_d$ is the daughter multiplicity.

The ATLAS $dE/dx$ signal gives a prediction for the gauge coupling $g_{Z'}$, up to a coefficient $|Q_q|\sqrt{B_d}$. The prediction is shown in table~\ref{tab:Lam}, under the simplifying assumption of a universal $U(1)'$ charge $Q_q$ for all quarks.\footnote{For non-universal quark charges, one can simply replace $Q_q$ with the weighted average $ \sum_i C_i Q_i / \sum_i C_i$, when it refers to the initial state, and with the average $ \sum_i Q_i / 5$, when it refers to the final state, where the sum extends over the first five quark species.}  As long as $B_d$ is not too small, the new gauge coupling constant $g_{Z'}$ is safely in the perturbative regime in the full range of relevant values of $\mpar$.

\begin{figure}[t]
\centering
\includegraphics[width=25em]{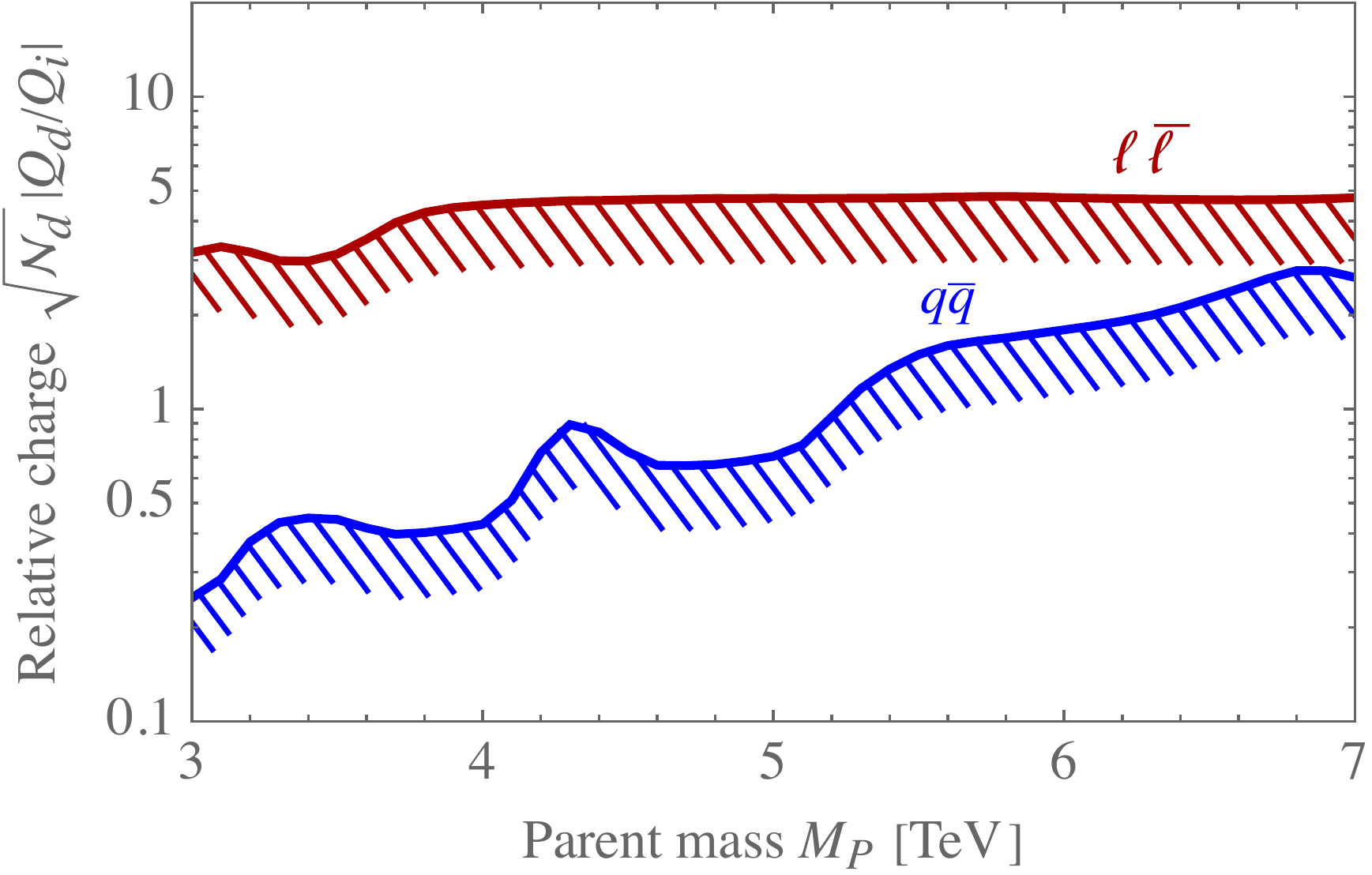} \caption{Lower bounds on the daughter $U(1)'$ charge in the combinations $\sqrt{{\cal N}_d}\, |Q_d/Q_q|$ (for the dijet channel) and $\sqrt{{\cal N}_d}\, |Q_d/Q_\ell |$ (for the dilepton channel) from CMS searches.
The value of the gauge coupling $g_{Z'}$ has been fixed to reproduce the ATLAS $dE/dx$ signal. 
\label{fig:dileptons}}
\end{figure}

As discussed at the beginning of this section, we expect an irreducible contribution to dijet events. Moreover, depending on $U(1)'$ charge assignments, we can also expect new effects in dilepton events, if $Z'$ has a significant decay width into leptons. Interestingly, once the value of $g_{Z'}$ is fixed to reproduce the ATLAS $dE/dx$ excess, the predictions for the dijet and dilepton cross sections are fully determined by $\mpar$ and the daughter effective charge $\sqrt{{\cal N}_d}\, Q_d$ in units of the quark charge $Q_q$ and lepton charge $Q_\ell$, respectively. The predictions are independent of $\Gamma_P$ and therefore are not affected by possible $Z'$ decay modes into other particles.

The present LHC limits on resonant contributions to dijet~\cite{CMS:2019gwf} and dilepton~\cite{CMS:2021ctt} events can be translated into lower bounds on $\sqrt{{\cal N}_d}\, |Q_d/Q_q|$  and $\sqrt{{\cal N}_d}\, |Q_d/Q_\ell |$, as shown in Fig.~\ref{fig:dileptons}. The figure shows that dijet limits are easily satisfied as long as the daughter charge is not much smaller than those of quarks. Dilepton searches provide stronger bounds on $Q_d$ with respect to the lepton charge. This may be taken as an indication that vector resonances with suppressed lepton couplings are favoured.

It is interesting to consider a $B$--$L$ gauge boson (such that $Q_q=1/3$ and $Q_\ell = -1$), where the predictions for dijet and dilepton resonant production are correlated. Figure~\ref{fig:dileptons} shows that dileptons are the most efficient channel to test a $B$--$L$ gauge boson. Present searches give the bound $\sqrt{{\cal N}_d}\, |Q_d| \gtrsim 5$, in the most relevant mass window $ \mpar \approx 4$--$6~{\rm TeV}$, and therefore require a sufficiently large daughter charge and/or multiplicity. We remark that the lower bound on $\sqrt{{\cal N}_d}\, |Q_d|$ scales as the inverse square root of the experimental efficiency in the $dE/dx$ signal, which we estimated as 20\%. A precise assessment of the bound would require a detailed experimental analysis and is quite sensitive to future statistical improvements.

Depending on $U(1)'$ charge assignment and the spectrum of new particles accompanying the daughters, other phenomenological signatures are possible. An intriguing example is the $Z'$ effect which could explain the recent $W$-mass anomaly claimed by the CDF collaboration~\cite{CDF:2022hxs}. Such an explanation requires a non-vanishing Higgs charge $Q_h$ under $U(1)'$, suppressed lepton charges, and a gauge coupling $g_{Z'}|Q_h| \simeq M_{Z'}/8 \TeV$~\cite{Strumia:2022qkt}. Therefore, the same $Z'$ can simultaneously fit both the $dE/dx$ excess and the $M_W$ measurement by CDF  if $\sqrt{B_d} Q_q/Q_h = \{ 0.035, 0.082,0.19,0.52\}$ for $M_Z'/{\rm TeV} = \{3,4,5,6 \}$. The emerging picture shows a $Z'$ coupled with comparable strength to quarks, Higgs and daughters, while couplings to leptons must be relatively suppressed. This might be indicative of a vector resonance of a coloured strongly-coupled sector in the multi-TeV range interacting with the Higgs boson, as in composite Higgs models.

\subsection*{EW-charged resonances}

In the cases discussed above, the parent resonance is assumed to be an $SU(2)_L$ singlet. However, a heavy electroweak doublet parent $P$ would in principle work as well. An important qualitative difference arising in this case is that $P$ cannot decay into a pair of the same daughter particle. As a consequence, scenarios with electroweak-charged resonances would be favoured if future data show that tracks with large $dE/dx$ are never accompanied by another ionising track from the recoiling particle.

In the case of an electroweak-doublet parent, a natural choice would be for it to decay into an $SU(2)_L$ triplet with hypercharge one (containing the electric charge-two state) and a doublet, possibly identified with the SM Higgs boson. Notice that the unit charge component of the triplet would give a lower $dE/dx \approx 1 \, \mathrm{MeV g^{-1} cm^2}$ signal, hidden in the large background. 

\subsection*{Coloured resonances}

Finally, we note that coloured parent resonances are more difficult to accommodate, if one insists to have decays into a pair of the same kind of daughter.  If the charge-two daughter were coloured it would have a significant QCD production with lower $\beta$, yielding a significant $dE/dx$ signal, close to the upper bound of the dynamic range of the ATLAS detector. Then the only possibility to explain the ATLAS excess is that the decay of the coloured resonance takes place into two different particles, and only the one with $Q \leq 1$ is coloured.

\section{Conclusions}
History has taught us to expect the unexpected in fundamental physics.  Not every discovery is foreseen, nor have they all provided the missing piece in an outstanding theoretical jigsaw puzzle.  This was true  for the archetypal LLP discovery of the muon.  In this discovery the muons were produced from the decays of heavier parent particles, the pions.  It is just a coincidence that the pion and muon masses are so close and, in principle, the parents could have been significantly heavier than the muons, boosting them in the parent rest-frame.

In this work we have considered whether history could repeat itself at the LHC by studying the phenomenology of boosted charged LLPs in $dE/dx$ searches.  We have shown that the phase space they occupy is distinct from commonly-considered scenarios where the LLPs are pair-produced in non-resonant processes. Plausible microscopic models of boosted LLPs exist, are consistent with present experimental limits, and can be searched for at future LHC runs. 
A general feature, independent of the specific model realisation of boosted LLPs, is an additional irreducible signature that could be revealed at resonant dijet searches, produced by the coupling of the parent resonance with SM light quarks or gluons. Other, more model-dependent, signatures can be useful to obtain further confirmation of potential discoveries.

An exciting aspect of boosted LLPs is that, so far, they are the only known explanation for the recently reported $dE/dx$ excess by the ATLAS collaboration, consistent with the information from the time-of-flight measurement, suggesting that $\beta \approx 1$. We find that overall the quality of the fit provided by boosted LLPs for the excess is very good and suggests new particles in the TeV range. It is also quite interesting that the excess lies in a low-background region and therefore it can turn into a more-than-$5\sigma$ discovery at the LHC Run 3, if present observations indicate
a real new-physics phenomenon.

Whether this excess will evolve into a full-blown `who ordered that?' discovery will be a question of statistics, a question of systematics, and ultimately a question of corroborating results from CMS.  Nevertheless, even if the excess eventually evaporates, heavy boosted charged LLPs will remain an interesting item on the menu of unexpected discoveries, which should be investigated by LHC experimental collaborations.

\section*{Acknowledgments}
We thank Ismet Siral for clarifications on~\cite{ATLAStalk} as well as the rest of the  ATLAS $dE/dx$ analysis group for comments on the manuscript.

\section*{Note added}
After this paper appeared as a preprint, the ATLAS collaboration presented new results in the search for long-lived multi-charged particles~\cite{ATLAS:2022cob}. Unlike our work, they only consider direct production from Drell-Yan or photon fusion, which gives rise to multi-charged \emph{unboosted} particles that cannot explain the $dE/dx$ excess. 
As we discussed in Sec.~\ref{sec:models}, this is however an irreducible complementary signal of our framework, for light-enough daughter particles. The ATLAS results for $Q=2$ exclude $m_D < 1.05\,\mathrm{TeV}$ at $95 \%$ C.L. for fermionic daughters while showing a mild excess, with  $4$ observed events in a region with $1.5$ expected background. For scalar daughters the corresponding limit inferred from the analysis in~\cite{Altakach:2022hgn} is $m_D \lesssim 700 \GeV$.

Recently, following previous hints~\cite{CMS:2019gwf, Dobrescu:2018psr}, the CMS collaboration has presented results for the dedicated analysis of heavy resonances decaying into two pairs of jets~\cite{CMS:2022usq}, a topology closely related to our explanation of the $dE/dx$ anomaly, excluding parent masses smaller than $7.6\TeV$ for a benchmark choice of couplings. The collaboration reports two anomalous events with parent mass $8$ TeV, decaying into a pair of $2$ TeV daughters. The former  could be potentially identified with our parent particle $P$, whereas the latter, being short-lived, cannot be our daughters $d$, but possibly EW partners of them.  



\section*{Appendix}

\appendix

\section{Calibration of the {\boldmath$dE/dx$} distribution}\label{app:calibration}
Here we give the results for the parameter extraction of the $dE/dx$ distribution, obtained by fitting the calibration data in~\cite{ATLAS:2022pib}. We find the best-fit values:
\begin{center}
\begin{tabular}{ccc|ccc}
$c_0$ & $c_1$ & $c_2$ & $\sigma$ & $\alpha$ & $n$ \\[0.2em]
\hline
\rule{0em}{1.2em} 0.81 & $-0.15$ & 0.20 & 0.162 & 1.34 & $\geq 8$
\end{tabular}
\end{center}
 The $c_i$  parameters enter the phenomenological MPV curve in eq.~\eqref{eq:dEdxMPV}, while the one-sided Crystal Ball distribution around it has Gaussian width $\sigma \,dE/dx|_{\rm MPV}$, and the $n$-th power-law starts at $\alpha \sigma \, dE/dx|_{\rm MPV}$ from the MPV. The parameters $c_i$ are given in units of ${\rm MeV\, g^{-1}\,cm^2}$.

\section{Validation of the analysis}\label{app:validation}
\begin{figure}[t]
\centering
\includegraphics[width=25em]{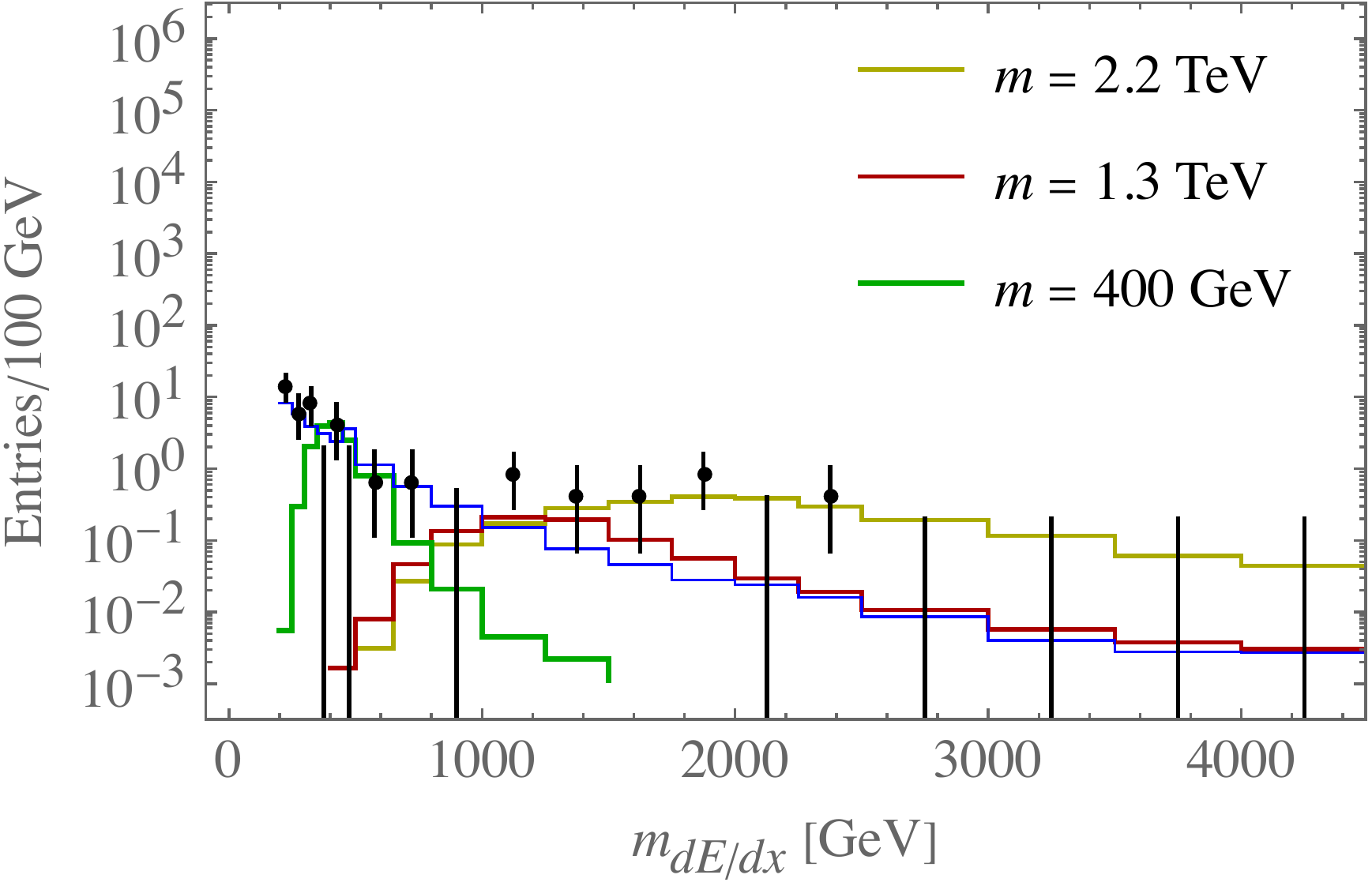} \caption{Signal models for the $Q=1$ hypothesis, to be compared with those reported in~\cite{ATLAS:2022pib}. For illustrative purposes, the overall signal strengths here are  chosen to match the benchmark models given in~\cite{ATLAS:2022pib}:  $2.2 \; \rm TeV$ gluinos (yellow line), $1.3\; \rm TeV$ charginos (red line) and $400\; \rm GeV$ sleptons (green line). We also show the distribution of the background (blue line).
\label{fig:validation}}
\end{figure}

In order to validate our simplified analysis against the one from ATLAS, we checked that we are able to reproduce the signal models in~\cite{ATLAS:2022pib} with sufficient accuracy. In particular, we performed a simulation analogous to the one described in the main body of the paper, but with $Q=1$.  Here we roughly approximated the $p_T$ distribution by a step function up to $\mpar$ and the $|\eta|$ distribution as a step function up to $1.2$. As shown in Fig.~\ref{fig:validation}, to be compared with the analogous one in~\cite{ATLAS:2022pib}, our simplified analysis is sufficient to reproduce the signal models given there. This gives us confidence that the dominant physical effects are captured by our simplified analysis.

\section{Details on the parameter fit}\label{app:fits} 
\begin{figure*}[t]
\centering
\includegraphics[width=20em]{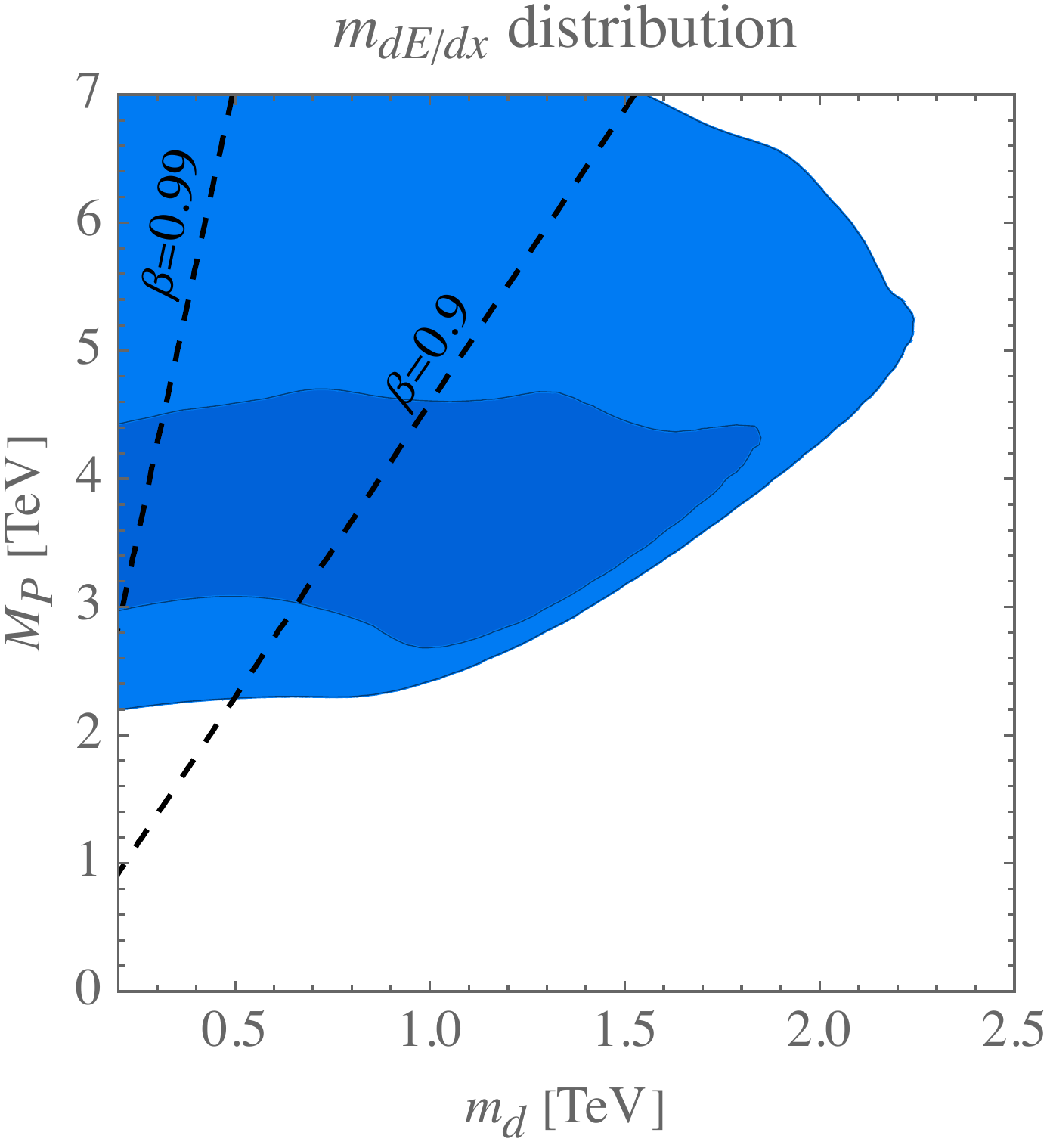} \caption{Profile-likelihood fits for the $m_{dE/dx}$ distribution only. Contours show the 1$\sigma$ and 2$\sigma$ preferred regions. The dashed lines denote the corresponding values of $\beta$ for a decay at rest of the parent resonance.
\label{fig:fit_m_only}}
\end{figure*}

In this appendix we give more details about the parameter fit discussed in Section~\ref{sec:results}. 

The simplest possibility would be to fit just the $m_{dE/dx}$ distribution, with results shown in Fig.~\ref{fig:fit_m_only}. However, this would be rather misleading, because only part of the available information is then used. In particular, while the regions $M_P \approx 2$--$3 \TeV$ and $\beta \lesssim 0.9$ look naively within the 2$\sigma$ favoured parameter space, the $p_T$ and $dE/dx$ distributions, respectively, are not properly reproduced.
The reason why $M_P \approx 2$--$3 \TeV$ fails to reproduce the $p_T$ distribution is that most of the excess in the $p_T$ histogram occurs for $p_T \gtrsim 750 \GeV$ (see Fig.~\ref{fig:benchmark_pT}), while the momentum of the reconstructed ionising particle is about $ \mpar/4$, a factor of two being due to the charge mismatch in the tracking reconstruction algorithm, as discussed in Sec.~\ref{sec:dEdx}. The reason why $\beta \lesssim 0.9$ fails to reproduce the $dE/dx$ distribution is manifest from Fig.~\ref{fig:beta}.

\begin{figure*}[t]
\centering
\includegraphics[width=25em]{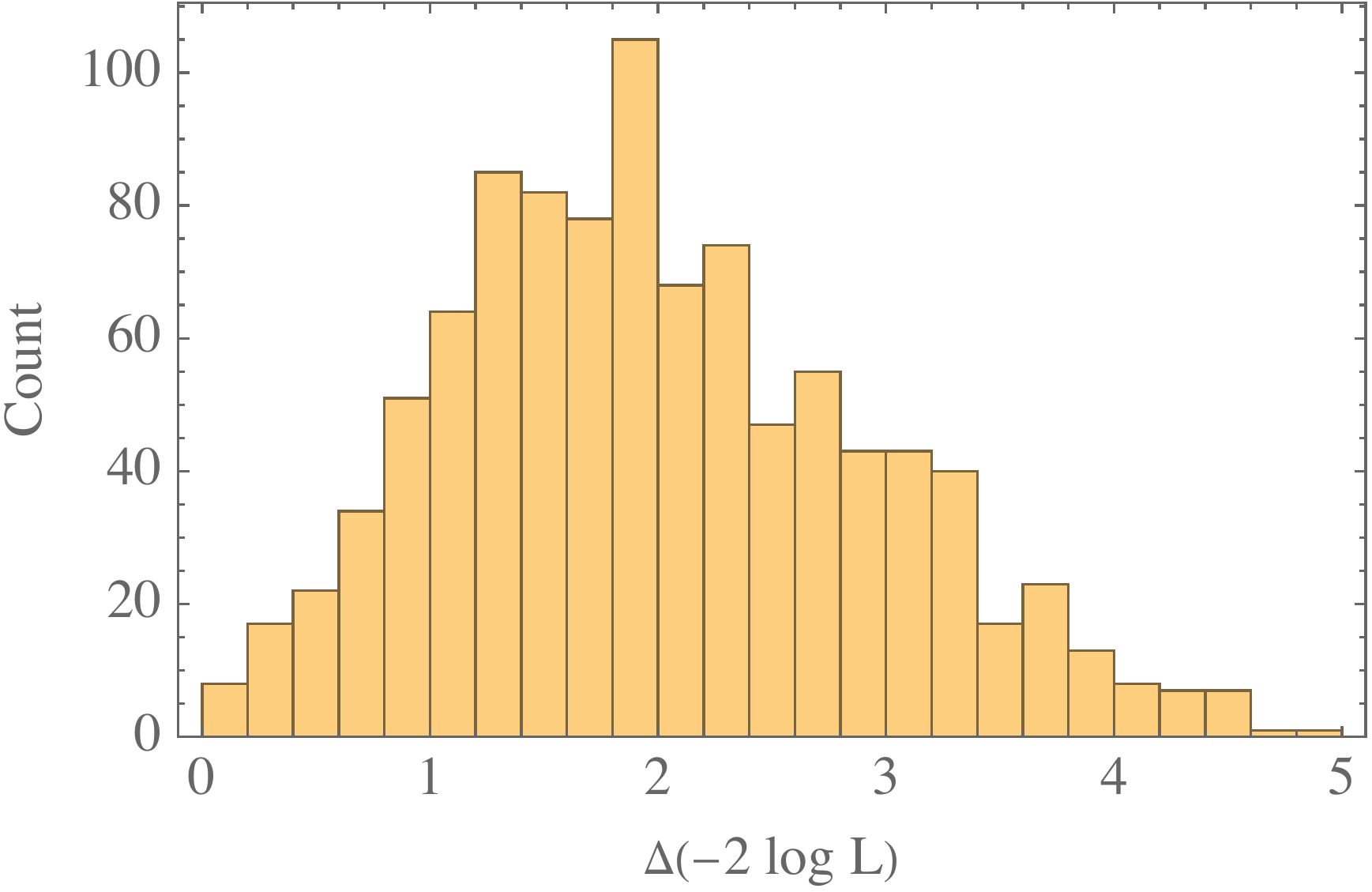} \caption{Distribution of the log-likelihood ratio in the toy pseudo-experiments.
\label{fig:toys}}
\end{figure*}

Therefore, in Fig.~\ref{fig:fit} we performed a combined fit of all three histograms. Given the correlations between them, the confidence intervals cannot be estimated by means of Wilks' theorem. Instead, we obtained them by means of toy pseudo-experiments, as follows. We approximate the confidence intervals as constant around the best-fit point $M_P \simeq 5.2 \TeV$, $m_d \simeq 650 \GeV$. For each pseudo-experiment, we assume new physics corresponding to the best-fit point and simulate a number of events Poisson-distributed around the best-fit value. We then build the toy-signal histograms, add them to the expected background, perform a toy fit, and calculate the log-likelihood ratio $\Delta(-2 \log L)$ with respect to the toy best-fit point. We run 1000 pseudo-experiments, obtaining the distribution of $\Delta(-2 \log L)$ plotted in Fig.~\ref{fig:toys}, which indeed is rather different from the would-be $\chi^2$ distribution with 2 degrees of freedom predicted by Wilks' theorem. From the distribution, we finally extract the 1$\sigma$ and 2$\sigma$ intervals as $\Delta(-2 \log L) = 2.4, 3.7$, respectively, which are used to generate Fig.~\ref{fig:fit}.

\bibliographystyle{JHEP}

\bibliography{refs}

\providecommand{\href}[2]{#2}\begingroup\raggedright\begin{thebibliography}{10}

\bibitem{Anderson:1936zz}
C.D.~Anderson and S.H.~Neddermeyer, \emph{{Cloud Chamber Observations of Cosmic
  Rays at 4300 Meters Elevation and Near Sea-Level}},
  \href{https://doi.org/10.1103/PhysRev.50.263}{\emph{Phys. Rev.} {\bfseries
  50} (1936) 263}.

\bibitem{Neddermeyer:1937md}
S.H.~Neddermeyer and C.D.~Anderson, \emph{{Note on the Nature of Cosmic Ray
  Particles}}, \href{https://doi.org/10.1103/PhysRev.51.884}{\emph{Phys. Rev.}
  {\bfseries 51} (1937) 884}.

\bibitem{Fairbairn:2006gg}
M.~Fairbairn, A.C.~Kraan, D.A.~Milstead, T.~Sjostrand, P.Z.~Skands and
  T.~Sloan, \emph{{Stable Massive Particles at Colliders}},
  \href{https://doi.org/10.1016/j.physrep.2006.10.002}{\emph{Phys. Rept.}
  {\bfseries 438} (2007) 1}
  [\href{https://arxiv.org/abs/hep-ph/0611040}{{\ttfamily hep-ph/0611040}}].

\bibitem{CMS:2011arq}
{\scshape CMS} collaboration, \emph{{Search for Heavy Stable Charged Particles
  in $pp$ collisions at $\sqrt{s}=7$ TeV}},
  \href{https://doi.org/10.1007/JHEP03(2011)024}{\emph{JHEP} {\bfseries 03}
  (2011) 024} [\href{https://arxiv.org/abs/1101.1645}{{\ttfamily 1101.1645}}].

\bibitem{ATLAS:2011ghv}
{\scshape ATLAS} collaboration, \emph{{Search for stable hadronising squarks
  and gluinos with the ATLAS experiment at the LHC}},
  \href{https://doi.org/10.1016/j.physletb.2011.05.010}{\emph{Phys. Lett. B}
  {\bfseries 701} (2011) 1} [\href{https://arxiv.org/abs/1103.1984}{{\ttfamily
  1103.1984}}].

\bibitem{ATLAS:2012urj}
{\scshape ATLAS} collaboration, \emph{{Searches for heavy long-lived sleptons
  and R-Hadrons with the ATLAS detector in $pp$ collisions at $\sqrt{s}=7$
  TeV}}, \href{https://doi.org/10.1016/j.physletb.2013.02.015}{\emph{Phys.
  Lett. B} {\bfseries 720} (2013) 277}
  [\href{https://arxiv.org/abs/1211.1597}{{\ttfamily 1211.1597}}].

\bibitem{CMS:2012wcg}
{\scshape CMS} collaboration, \emph{{Search for heavy long-lived charged
  particles in $pp$ collisions at $\sqrt{s}=7$ TeV}},
  \href{https://doi.org/10.1016/j.physletb.2012.06.023}{\emph{Phys. Lett. B}
  {\bfseries 713} (2012) 408}
  [\href{https://arxiv.org/abs/1205.0272}{{\ttfamily 1205.0272}}].

\bibitem{ATLAS:2013cab}
{\scshape ATLAS} collaboration, \emph{{Search for long-lived, multi-charged
  particles in pp collisions at $\sqrt{s}$=7 TeV using the ATLAS detector}},
  \href{https://doi.org/10.1016/j.physletb.2013.04.036}{\emph{Phys. Lett. B}
  {\bfseries 722} (2013) 305}
  [\href{https://arxiv.org/abs/1301.5272}{{\ttfamily 1301.5272}}].

\bibitem{CMS:2013czn}
{\scshape CMS} collaboration, \emph{{Searches for Long-Lived Charged Particles
  in $pp$ Collisions at $\sqrt{s}$=7 and 8 TeV}},
  \href{https://doi.org/10.1007/JHEP07(2013)122}{\emph{JHEP} {\bfseries 07}
  (2013) 122} [\href{https://arxiv.org/abs/1305.0491}{{\ttfamily 1305.0491}}].

\bibitem{ATLAS:2015wsk}
{\scshape ATLAS} collaboration, \emph{{Search for metastable heavy charged
  particles with large ionisation energy loss in pp collisions at $\sqrt{s} =
  8$ TeV using the ATLAS experiment}},
  \href{https://doi.org/10.1140/epjc/s10052-015-3609-0}{\emph{Eur. Phys. J. C}
  {\bfseries 75} (2015) 407}
  [\href{https://arxiv.org/abs/1506.05332}{{\ttfamily 1506.05332}}].

\bibitem{ATLAS:2016tbt}
{\scshape ATLAS} collaboration, \emph{{Search for metastable heavy charged
  particles with large ionization energy loss in pp collisions at $\sqrt{s} =
  13$ TeV using the ATLAS experiment}},
  \href{https://doi.org/10.1103/PhysRevD.93.112015}{\emph{Phys. Rev. D}
  {\bfseries 93} (2016) 112015}
  [\href{https://arxiv.org/abs/1604.04520}{{\ttfamily 1604.04520}}].

\bibitem{CMS:2016kce}
{\scshape CMS} collaboration, \emph{{Search for long-lived charged particles in
  proton-proton collisions at $\sqrt s=$ 13 TeV}},
  \href{https://doi.org/10.1103/PhysRevD.94.112004}{\emph{Phys. Rev. D}
  {\bfseries 94} (2016) 112004}
  [\href{https://arxiv.org/abs/1609.08382}{{\ttfamily 1609.08382}}].

\bibitem{PDG}
{\scshape Particle Data Group} collaboration, \emph{{Review of Particle
  Physics}}, \href{https://doi.org/10.1093/ptep/ptaa104}{\emph{PTEP} {\bfseries
  2020} (2020) 083C01}.

\bibitem{ATLAS:2022pib}
{\scshape ATLAS} collaboration, \emph{{Search for heavy, long-lived, charged
  particles with large ionisation energy loss in $pp$ collisions at $\sqrt{s} =
  13~\text{TeV}$ using the ATLAS experiment and the full Run 2 dataset}},
  \href{https://arxiv.org/abs/2205.06013}{{\ttfamily 2205.06013}}.

\bibitem{ATLAStalk}
{\scshape ATLAS} collaboration, ``{\it Search for heavy, long-lived, charged
  particles with large ionisation energy loss in $pp$ collisions at $\sqrt{s} =
  13$ {TeV} using the {ATLAS} experiment and the full Run 2 dataset}.''
  \href{https://agenda.infn.it/event/28365/contributions/161449/attachments/89009/119418/LeThuile-dEdx.pdf}{Presentation}
  by Ismet Siral at the La Thuile 2022 Conference.

\bibitem{Liang:2014via}
{\scshape ATLAS} collaboration, \emph{{Performance and operation experience of
  the Atlas Semiconductor Tracker}},
  \href{https://doi.org/10.1142/S2010194514602956}{\emph{Int. J. Mod. Phys.
  Conf. Ser.} {\bfseries 31} (2014) 1460295}
  [\href{https://arxiv.org/abs/1402.6506}{{\ttfamily 1402.6506}}].

\bibitem{Benekos:2003xra}
N.C.~Benekos, R.~Clifft, M.~Elsing and A.~Poppleton, ``{ATLAS Inner Detector
  Performance}.'' \href{http://cds.cern.ch/record/688762}{ATL-INDET-2004-002},
  ATL-COM-INDET-2003-023.

\bibitem{Akhmedov:2021qmr}
E.~Akhmedov, \emph{{Nuclear fusion catalyzed by doubly charged scalars:
  Implications for energy production}},
  \href{https://arxiv.org/abs/2109.13960}{{\ttfamily 2109.13960}}.

\bibitem{Martin:2009iq}
A.D.~Martin, W.J.~Stirling, R.S.~Thorne and G.~Watt, \emph{{Parton
  distributions for the LHC}},
  \href{https://doi.org/10.1140/epjc/s10052-009-1072-5}{\emph{Eur. Phys. J. C}
  {\bfseries 63} (2009) 189} [\href{https://arxiv.org/abs/0901.0002}{{\ttfamily
  0901.0002}}].

\bibitem{CMS:2019gwf}
{\scshape CMS} collaboration, \emph{{Search for high mass dijet resonances with
  a new background prediction method in proton-proton collisions at $\sqrt{s}
  =$ 13 TeV}}, \href{https://doi.org/10.1007/JHEP05(2020)033}{\emph{JHEP}
  {\bfseries 05} (2020) 033}
  [\href{https://arxiv.org/abs/1911.03947}{{\ttfamily 1911.03947}}].

\bibitem{Bauer:2009cc}
C.W.~Bauer, Z.~Ligeti, M.~Schmaltz, J.~Thaler and D.G.E.~Walker,
  \emph{{Supermodels for early LHC}},
  \href{https://doi.org/10.1016/j.physletb.2010.05.032}{\emph{Phys. Lett. B}
  {\bfseries 690} (2010) 280}
  [\href{https://arxiv.org/abs/0909.5213}{{\ttfamily 0909.5213}}].

\bibitem{CMS:2021ctt}
{\scshape CMS} collaboration, \emph{{Search for resonant and nonresonant new
  phenomena in high-mass dilepton final states at $ \sqrt{s} $ = 13 TeV}},
  \href{https://doi.org/10.1007/JHEP07(2021)208}{\emph{JHEP} {\bfseries 07}
  (2021) 208} [\href{https://arxiv.org/abs/2103.02708}{{\ttfamily
  2103.02708}}].

\bibitem{CDF:2022hxs}
{\scshape CDF} collaboration, \emph{{High-precision measurement of the $W$
  boson mass with the CDF II detector}},
  \href{https://doi.org/10.1126/science.abk1781}{\emph{Science} {\bfseries 376}
  (2022) 170}.

\bibitem{Strumia:2022qkt}
A.~Strumia, \emph{{Interpreting electroweak precision data including the
  $W$-mass CDF anomaly}},  \href{https://arxiv.org/abs/2204.04191}{{\ttfamily
  2204.04191}}.

\bibitem{ATLAS:2022cob}
{\scshape ATLAS} collaboration, ``{Search for heavy long-lived multi-charged
  particles in the full Run-II $pp$ collision data at $\sqrt{s}$ = 13 TeV using
  the ATLAS detector}.''
  \href{https://inspirehep.net/literature/2095815}{ATLAS-CONF-2022-034}, 2022.

\bibitem{Altakach:2022hgn}
M.M.~Altakach, P.~Lamba, R.~Mase\l{}ek, V.A.~Mitsou and K.~Sakurai,
  \emph{{Discovery prospects for long-lived multiply charged particles at the
  LHC}},  \href{https://arxiv.org/abs/2204.03667}{{\ttfamily 2204.03667}}.

\bibitem{Dobrescu:2018psr}
B.A.~Dobrescu, R.M.~Harris and J.~Isaacson, \emph{{Ultraheavy resonances at the
  LHC: beyond the QCD background}},
  \href{https://arxiv.org/abs/1810.09429}{{\ttfamily 1810.09429}}.

\bibitem{CMS:2022usq}
{\scshape CMS} collaboration, \emph{{Search for resonant and nonresonant
  production of pairs of dijet resonances in proton-proton collisions at
  $\sqrt{s}$ = 13 TeV}},  \href{https://arxiv.org/abs/2206.09997}{{\ttfamily
  2206.09997}}.

\end{thebibliography}\endgroup

\end{document}